\documentclass[namedreferences]{solarphysics}

\usepackage[hyperref,optionalrh,showbiblabels]{spr-sola-addons} 
\usepackage{graphicx}        
\usepackage{amssymb}        
\usepackage{color}           

\begin{document}
\begin{article}
\begin{opening}
\title{New Homogeneous Dataset of Solar EUV synoptic maps from SOHO/EIT and SDO/AIA\\ {\it Solar Physics}}

\author[addressref={aff1},corref ,email={amr.hamada@oulu.fi}]{\fnm{A.}~\lnm{Hamada$^{\ast}$}}
\author[addressref={aff1},email={timo.Asikainen@oulu.fi}]{\fnm{T.}~\lnm{Asikainen}}
\author[addressref={aff1},email={Kalevi.Mursula@oulu.fi}]{\fnm{K.}~\lnm{Mursula}}


\address[id=aff1]{ReSoLVE Centre of Excellence, Space Climate Research Unit, University of Oulu, Finland}

\runningauthor{A.Hamada et al.}
\runningtitle{Example paper}

\begin{abstract}
Synoptic maps of solar EUV intensities have been constructed for many decades in order to display the distribution of the different EUV emissions across the solar surface, with each map representing one Carrington rotation (i.e., one rotation of the Sun). This paper presents a new solar EUV synoptic map dataset based on full-disk images from {\it Solar and Heliospheric Observatory/Extreme Ultraviolet Imaging Telescope} (SOHO/EIT) and {\it Solar Dynamics Observatory/Atmospheric Imaging Assembly} (SDO/AIA). In order to remove the significant and complicated drift of EIT and AIA EUV intensities due to sensor degradation, we construct the synoptic maps in standardized intensity scale. We describe a method of homogenizing the SOHO/EIT maps with SDO/AIA maps by transforming the EIT intensity histograms to AIA level. The new maps cover the years from 1996 to 2018 with 307 SOHO/EIT and 116 SDO/AIA synoptic maps, respectively. These maps provide a systematic and homogenous view of the entire solar surface in four EUV wavelengths, and are well suited, e.g., for studying long-term coronal hole evolution.

\end{abstract}

\end{opening}

\section{Introduction}
     \label{S-Introduction} 

Synoptic maps provide a convenient way to display the evolution/distribution of various physical quantities and features on the entire solar surface. For example, synoptic maps have been used to study coronal holes (CH) \citep{Karna2015, Golubeva2017, Hamada2018}, photospheric and coronal magnetic fields \citep{Virtanen2016} and long-lived active regions (ARs). Solar synoptic maps can also be used to study the nonuniform structure of the EUV corona in both longitude and latitude, reflecting the nonuniform distribution of the large-scale magnetic field \citep{Benevolenskaya2001}. Different ground-based observations of the photospheric magnetic field are also presented in the form of synoptic maps (e.g., Wilcox Solar Observatory, National Solar Observatory, Mount Wilson Observatory and  Kitt Peak Vacuum Telescope).\\
 
Synoptic maps are constructed by concatenating a series of  meridian strips taken from full-disk images covering a full Carrington Rotation (CR) of 27.2753 consecutive days. Each individual full-disk image is remapped into heliographic coordinates with longitude along the x-axis, and latitude (or sine-latitude) along the y-axis. A rigid solar rotation rate is often assumed in order to reduce the effects of the dynamic surface of the Sun to the synoptic maps \citep{harvey1998}.\\

In this study, we present a new dataset of EUV synoptic maps, based on {\it Extreme Ultraviolet Imaging Telescope} (EIT) and {\it Atmospheric Imaging Assembly} (AIA) observations on board of {\it Solar and Heliospheric Observatory} (SOHO) and {\it Solar Dynamics Observatory} (SDO), respectively. These maps are constructed homogeneously with the same methodology and daily resolution (13.3$^\circ$ wide central solar meridian strip) for both instruments and finally inter-calibrated to eliminate the differences in intensity distribution between the two instruments. Moreover, SDO/AIA data are used to construct also EUV synoptic maps with much more narrow meridian strips (1.1$^\circ$ wide) representing a closer match between time and longitude in synoptic maps.

This paper is organized as follows. Section~\ref{S-Previous_dataset} presents techniques previously used to construct EUV synoptic maps from SOHO/EIT and SDO/AIA images. Section~\ref{S-data_analysis} discusses EIT and AIA full-disk images, presenting the filtration and calibration criteria used to avoid corrupted images and to correct certain other problems. Section~\ref{S-synop_maps} discusses the procedure for constructing the synoptic maps, including projection to heliographic coordinates, stripping, and merging of strips into a continuous synoptic map. The new dataset is compared with other publicly available datasets, and improvements are discussed. Section~\ref{S-eit_aia} discusses the homogenization of EIT maps with AIA maps, providing a continuous and uniform dataset. Section~\ref{S-Summary} summarizes our results.

\section{Previous EUV synoptic map datasets}
\label{S-Previous_dataset}

SOHO/EIT instrument \citep{Delaboudiniere1995} takes images of the Sun in four passbands that are centered on intense emission lines of the solar EUV spectrum. He II (304{\AA}), Fe XI/Fe X (171{\AA}), Fe XII (195{\AA}), and Fe XV (284{\AA}) lines provide temperature diagnostics in the range from 6 $\times$  $10^4\,{\rm K}$ to 3 $\times$  $10^6\,{\rm K}$. These different wavelengths represent solar layers at different altitudes. He II 304{\AA} images are dominated by emissions from structures of the transition region network \citep{Benevolenskaya2001}, indicating the magnetic footpoints of coronal loops and outlining the bases of coronal holes. Fe XI/Fe X 171{\AA} images display background emission that is present over most of the quiet Sun (QS), while the most intense emission comes from active regions with closed magnetic field. Fe XII 195{\AA} images are also dominated by emission from the closed magnetic field regions of the Sun, showing the inner solar corona with different distributions of intensities for CHs and the QS. The Fe XV 284{\AA} passband allows an analysis of the hotter active regions. 

For SDO/AIA instrument \citep{Lemen2012}, solar images are taken at 7 wavelengths in EUV (304{\AA}, 171{\AA}, 193{\AA}, 211{\AA}, 335{\AA}, 94{\AA}, 131{\AA}) and 3 in UV/visible (white light, 1700{\AA}, 1600{\AA}). Only four EUV channels (304{\AA}, 171{\AA}, 193{\AA} and 211{\AA}) are selected here in order to be compatible with EIT. Table \ref{T1} summarizes the characteristics of selected EIT and AIA wavelengths \citep{Moses1997,Petkaki_2012}. Although there are subtle differences in the temperature response functions of the AIA/193{\AA} and EIT/195{\AA} channels, they are similar enough to be used together to generate synchronic maps \citep{Caplan2016}.
The EUV emissions of AIA/211{\AA} and EIT/284{\AA} originate from different ions in the corona, Fe-XIV and Fe-XV, respectively.
However, the temperature responses of these two spectral lines have a large overlap and both peak close to each other at about 2.0 MK temperature \citep{Moses1997,Petkaki_2012}.
Therefore it is unlikely that there are large systematic differences between the AIA/211{\AA} and EIT/284{\AA} images.

\begin{table}
\caption{SOHO/EIT and SDO/AIA wavelengths}
\label{T1}
\begin{tabular}{cccc}     
\hline               
  Source & Wavelength & Ion & Peak Temperature (MK)\\
\hline               
SOHO/EIT & 284{\AA}  & Fe XV& ~2.0 \\
SDO/AIA & 211{\AA}  & Fe XIV & ~2.0 \\
SOHO/EIT & 195{\AA}  & Fe XII & ~1.6 \\
SDO/AIA & 193{\AA}  & Fe XII & ~1.6 \\
SOHO/EIT and SDO/AIA & 171{\AA}  & Fe IX-X & ~1.3 \\
SOHO/EIT and SDO/AIA & 304{\AA}  & He II & ~0.08 \\               
\hline               
\end{tabular}
\end{table}

Currently, there are two incomplete EUV synoptic map datasets based on SOHO/EIT observations and one synoptic map dataset based on SDO/AIA observations. \cite{Benevolenskaya2001} constructed SOHO/EIT synoptic maps in three Fe (171{\AA}, 195{\AA}, 284{\AA}) lines and He II (304{\AA}) from CR 1911 (1996 June 28) to CR 2055 (2007 March 31) by concatenating 16$^\circ$ wide central meridian strips. The size of these synoptic maps is 167 $\times$ 360 pixels with a resolution of $1^\circ \times 1^\circ$ of heliographic latitude and longitude covering all longitudes (1$^\circ$ to 360$^\circ$) and most latitudes ($-83^\circ$ to $+83^\circ$, leaving out those polar latitudes that are not always visible). These EIT synoptic maps are provided by Stanford Solar Observatories Group (SSOG) (\url{http://sun.stanford.edu/synop/EIT/index.html}). Although for every synoptic map, both sine and linear latitude GIF images were presented, the files offered in FITS (Flexible Image Transport System) format are in linear latitude only. Also, the maps from CR 1911 (1996 June 28) to CR 2042 (2006 April 10) are based on calibrated data, and from CR 2043 (2006 May 08) to CR 2055 (2007 March 31) on preliminary calibrated data.\\

Another set of EIT synoptic maps for only three wavelengths (171{\AA}, 195{\AA} and 304{\AA}) are provided by the Space Weather Lab (SWL) at George Mason University (\url{http://spaceweather.gmu.edu/projects/synop/EITSM.html})  \citep{HessWebber2014}. The available maps extend from CR 2058 (2007 June 21) to CR 2102 (2010 October 03). Thus, the SSOG maps and the SWL maps are disparate and have a gap of two Carrington rotations in-between. In the SWL maps, central meridian longitudinal strips of 13.63$^\circ$ width have been concatenated using four images per day with 3/4 overlap between adjacent strips. The size of these synoptic maps is 3600$\times$1080 pixels ($0.1^\circ\times 0.1667^\circ$) and cover all longitudes and all measured latitudes at best up to $\pm90^\circ$. There are many missing maps in this dataset (23 out of 45 CRs) but the corresponding FITS files exist in both sine and linear latitude.\\

\begin{figure} 
\centerline{\includegraphics[width=1.0\textwidth,clip=]{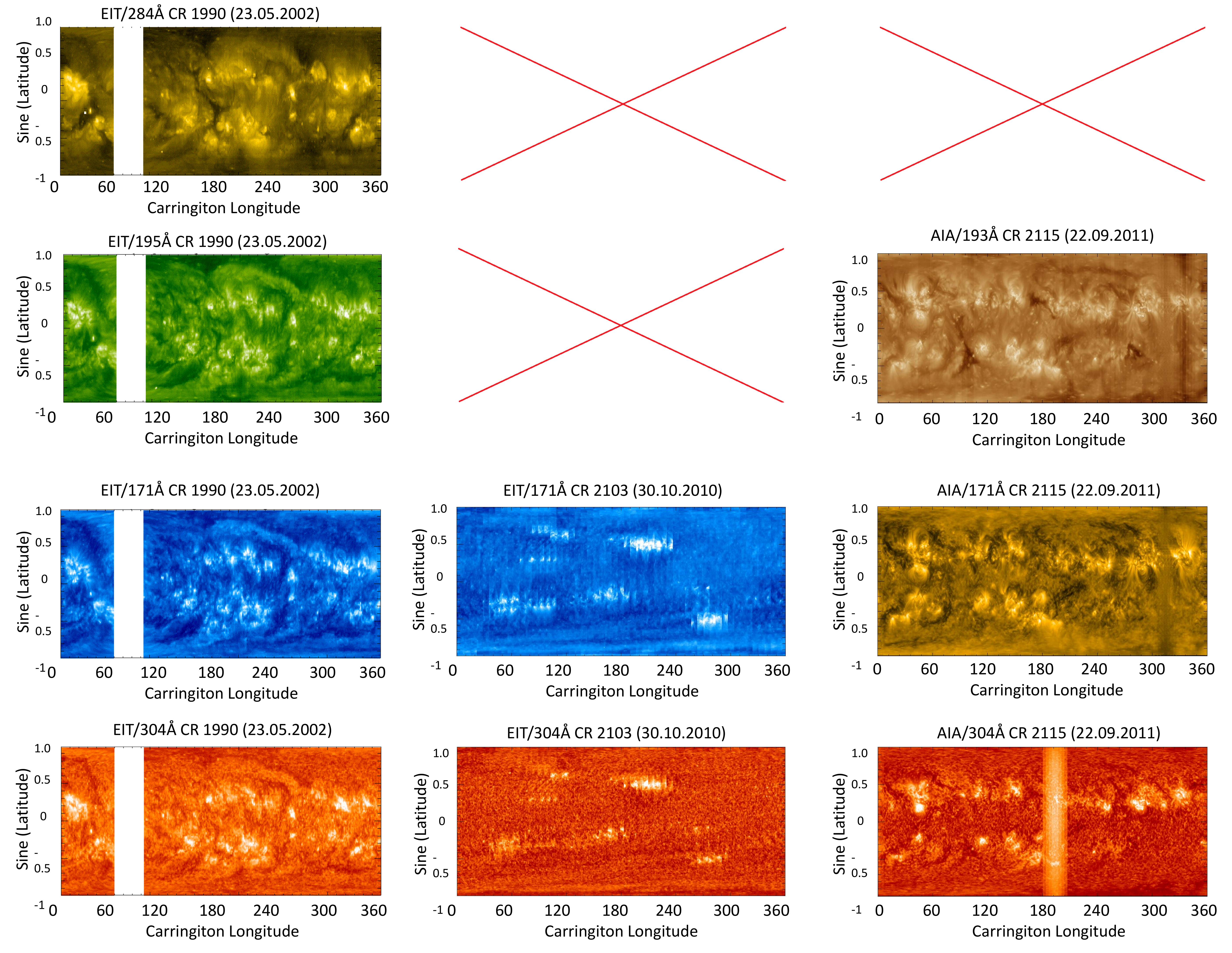}}
\caption{Examples of EUV synoptic maps for SOHO/EIT and SDO/AIA. First column: SOHO/EIT maps from Stanford Solar Observatories Group for CR 1990 (2002 May 23) in four wavelengths (284{\AA}, 195{\AA}, 171{\AA} and 304{\AA}). Second column: SOHO/EIT maps from Space Weather Lab in two wavelengths (171{\AA} and 304{\AA}; 195{\AA} missing) for CR 2103 (2010 October 30). Third column: SDO/AIA maps from Space Weather Lab in three wavelengths (193{\AA}, 171{\AA} and 304{\AA}) for CR 2115 (2011 September 22).}\label{F1}
\end{figure}

The only dataset of SDO/AIA synoptic maps is provided by SWL (\url{http://spaceweather.gmu.edu/projects/synop/AIASM.html}) from CR 2097 (2010 May 20) to CR 2186 (2017 January 10) \citep{Karna2014}. The resolution and latitudinal extent is the same as that of SWL/EIT synoptic maps (13.63$^\circ$ wide central meridian strips). Also, the maps are provided only for three wavelengths (171{\AA}, 193{\AA} and 304{\AA}). The maps from CR 2097 (2010 May 20) to CR 2124 (2012 May 25) are available in sine and linear latitude while from CR 2125 (2012 June 21) to CR 2186 (2017 January 10) the maps are available in linear latitude only. The temporal coverage of all the three databases of SOHO/EIT and SDO/AIA synoptic maps are summarized in Table~\ref{T2}. Figure \ref{F1} shows a sample of synoptic maps for the four EUV wavelengths from the three datasets. Many of the maps contain a data gap (represented as a white space in SSOG/EIT maps in Fig. \ref{F1}, first column). Also, sharp edges between adjacent longitudinal strips are observed in most of the maps from all datasets (Fig. \ref{F1}, second column). Maps of some wavelengths are missing for some specific Carrington rotations. For example, EIT/195{\AA} map is missing for CR 2103 in SWL/EIT maps (Figure \ref{F1}, second row; second column). For SWL/AIA maps, it also seems that the camera's exposure time and the image quality key-factors are not taken into account in the map preparation which is sometimes seen as overexposed (or underexposed) strips in the synoptic maps (for an example see Figure \ref{F1}, 3rd column, bottom row).\\
 
While inspecting these synoptic maps, we \citep{Hamada2018} have earlier noted on the degradation of the AIA instrument with time and on differences in pixel intensity histograms due to different sensor responses of EIT and AIA. We also found that the SWL maps erroneously contained the RGB values of pixels instead of line intensity. This error has been corrected in the SWL dataset since then (N. Karna, 2017, private communication). In a subsequent, more detailed inspection, we have found, as noted above, that sharp edges often appear between adjacent strips in most EIT and AIA synoptic maps. 
 
\begin{table}
\caption{Data intervals of previous SOHO/EIT and SDO/AIA EUV synoptic maps}
\label{T2}
\begin{tabular}{cccccc}     
\hline               
  Source & Observable & \multicolumn{2}{c}{Start} & \multicolumn{2}{c}{End} \\
 &  & CR & Date & CR & Date\\
\hline               
SSOG& SOHO/EIT: 284{\AA}, 195{\AA}, 171{\AA}, 304{\AA}  & 1911 & ~1996.06.28 & 2055 &  ~2007.03.31 \\
SWL&   SOHO/EIT: 195{\AA}, 171{\AA}, 304{\AA}  &  2058 &  ~2007.06.21 &  2102 &  ~2010.10.03 \\
SWL&  SDO/AIA: 193{\AA}, 171{\AA}, 304{\AA}    &  2097 &  ~2010.05.20 &  2186 &  ~2017.01.10 \\            
\hline               
\end{tabular}
\end{table}

In this paper, we use SOHO/EIT and SDO/AIA full-disk images to construct a new dataset of synoptic maps using strips of same width (13.3$^\circ$) for both EIT and AIA.
Additionally we construct AIA maps also with narrower 1.1$^\circ$ strips which allow better correspondence of synoptic map longitudes with the central solar meridian.

\section{Data analysis} 
      \label{S-data_analysis}      
Full-disk solar images from both SOHO/EIT and SDO/AIA are used to construct the new EUV synoptic maps. SOHO/EIT images are available from January 1996 until the end of December 2018 while SDO/AIA images are available from May 2010 until present. 
Table \ref{T3} shows SOHO/EIT and SDO/AIA full-disk data intervals used in this study. Each full-disk image was 
automatically checked to avoid any corrupted/noisy data. \\

\begin{table}
\caption{Data intervals of SOHO/EIT and SDO/AIA full disk observations used in this study}
\label{T3}
\begin{tabular}{cccc}     
\hline               
  Source & Observable & Start & End \\
\hline               
SOHO/EIT&  284{\AA}, 195{\AA}, 171{\AA} and 304{\AA}  &  ~1996.01.16 &  ~2018.12.31 \\
SDO/AIA&  211{\AA}, 193{\AA}, 171{\AA} and 304{\AA}  &  ~2010.05.13 &  ~2018.12.31 \\
\hline               
\end{tabular}
\end{table}

\begin{figure} 
\centerline{\includegraphics[width=1.0\textwidth,clip=]{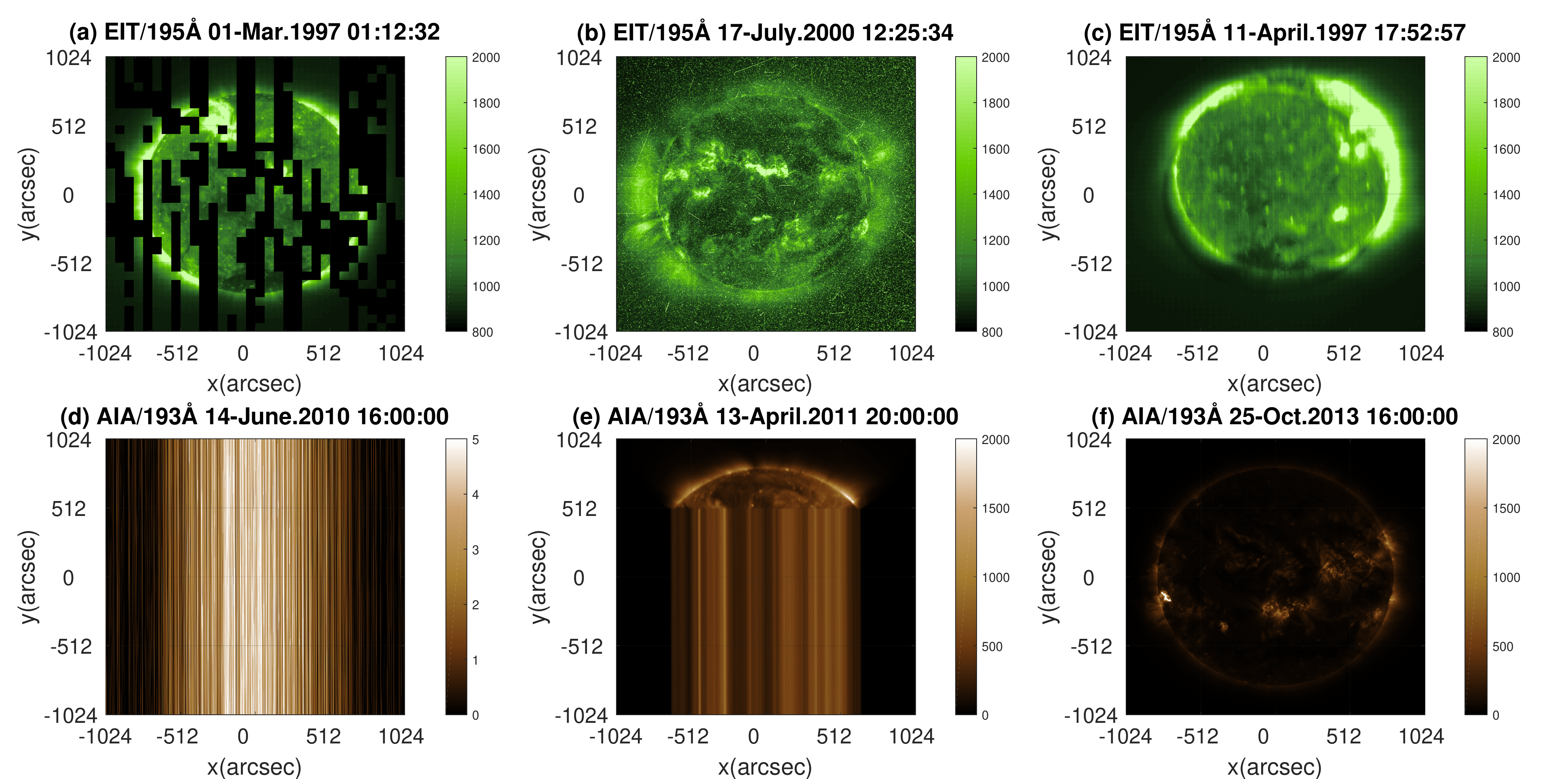}}
\caption{Examples of corrupted images that can be found within the SOHO/EIT (a, b and c) and SDO/AIA (d, e and f) datasets. (a) EIT image with a number of missing blocks identified by \MakeUppercase{N\_MISSING\_BLOCKS} keyword. (b) High noise EIT image due to solar energitic partcles. (c) Poor quality doubled exposure EIT image. (d, e) Two corrupted AIA images identified by \MakeUppercase{QUALITY} keyword. (f)An underexposed AIA image identified by \MakeUppercase{EXP\_TIME} keyword.}\label{F2}
\end{figure}

\subsection{SOHO/EIT} 
  \label{S-soho_eit}

SOHO/EIT full-disk images were downloaded from the SOHO Science Archive (SSA) hosted at the European Space Astronomy Centre (ESAC), and National Aeronautics and Space Administration (NASA).  SOHO has an extensive database including over 20 years of data from 1996 to 2018 (Table ~\ref{T3}). EIT routinely takes one image in each of its four wavelengths several times per day. Each image is 1024 $\times$1024 pixels, corresponding to a resolution of 2.6 arcsec per pixel. 

We used different keywords to identify different types of image deficiencies and noise. 
Keyword \MakeUppercase{N\_MISSING\_BLOCKS} represents the number of missing blocks in the EIT images due to, e.g., unstable EIT telemetry (for an example see Figure \ref{F2}a). Another reason for poor image quality are solar energetic particles, which can greatly increase the image noise (for an example see Figure \ref{F2}b). We used the mean intensity of the off-limb area as a measure to eliminate the noisy images. In addition to missing blocks and noise the EIT images may be subject to a higher level of scattered light \citep{Scherrer2012}. All images with enhanced noise or corrupted/missing pixel blocks are discarded and only good quality images were processed. Keyword  \MakeUppercase{EXPTIME} was used to avoid all images with very high or low exposure times (Figure ~\ref{F2}c). \MakeUppercase{SC\_ROLL} keyword represents the roll angle offsets from the projection of the Sun's North Pole and it is used to ensure that the solar north is at top of the image. \\

\begin{figure} 
\centerline{\includegraphics[width=1.0\textwidth,clip=]{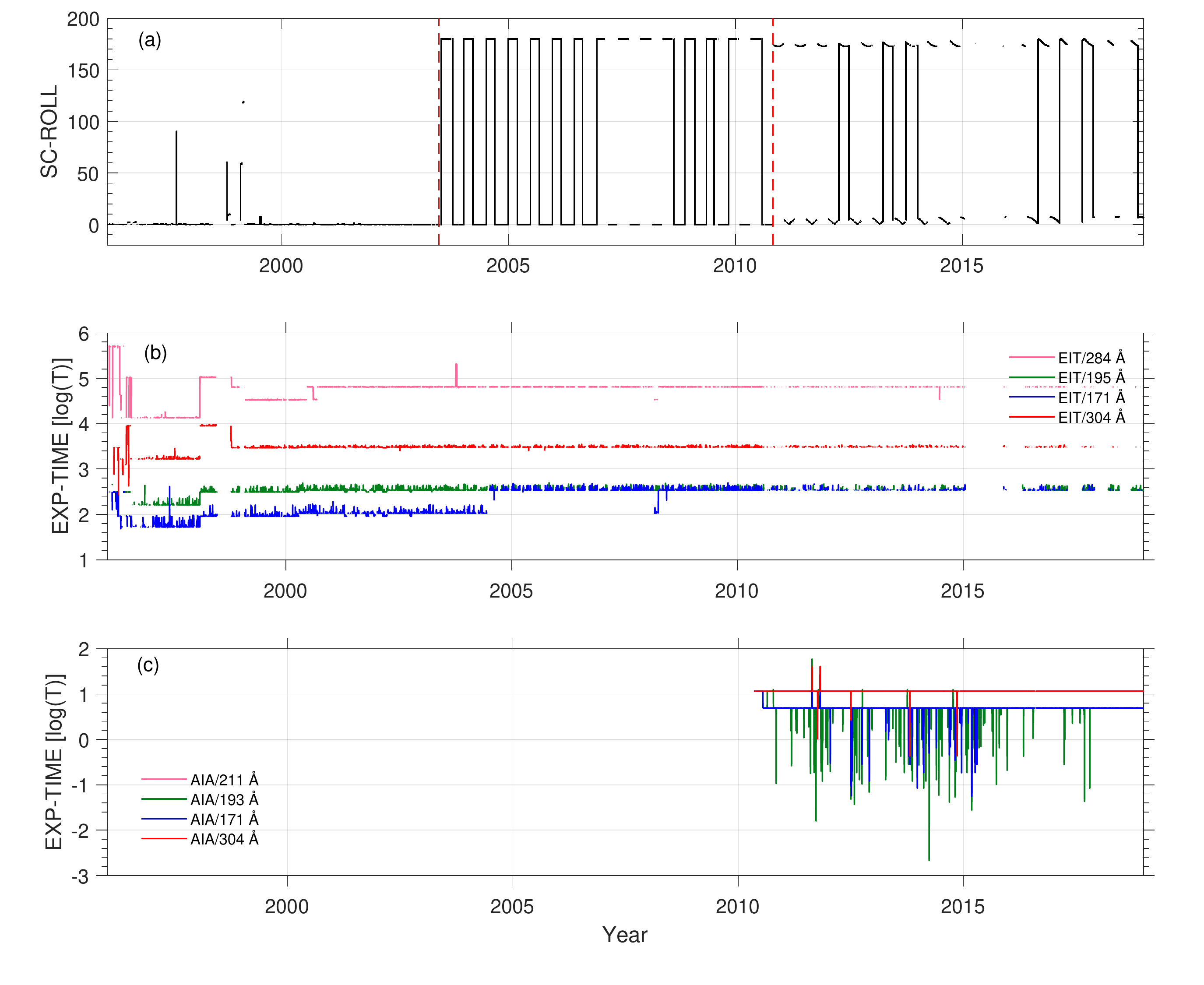}}
\caption{(a) SOHO/EIT's CCD logarithmic exposure time 284{\AA}, 195{\AA}, 171{\AA} and 304{\AA} different spectral lines. (b) Spacecraft roll angle in degrees counter-clockwise. Vertical dashed red lines mark the change in the spacecraft orientation at June 19, 2003  and October 29, 2010. (c) SDO/AIA's CCD logarithmic exposure time at 211{\AA}, 193{\AA}, 171{\AA} and 304{\AA} spectral lines. }\label{F3}
\end{figure}

\begin{figure} 
\centerline{\includegraphics[width=1.0\textwidth,clip=]{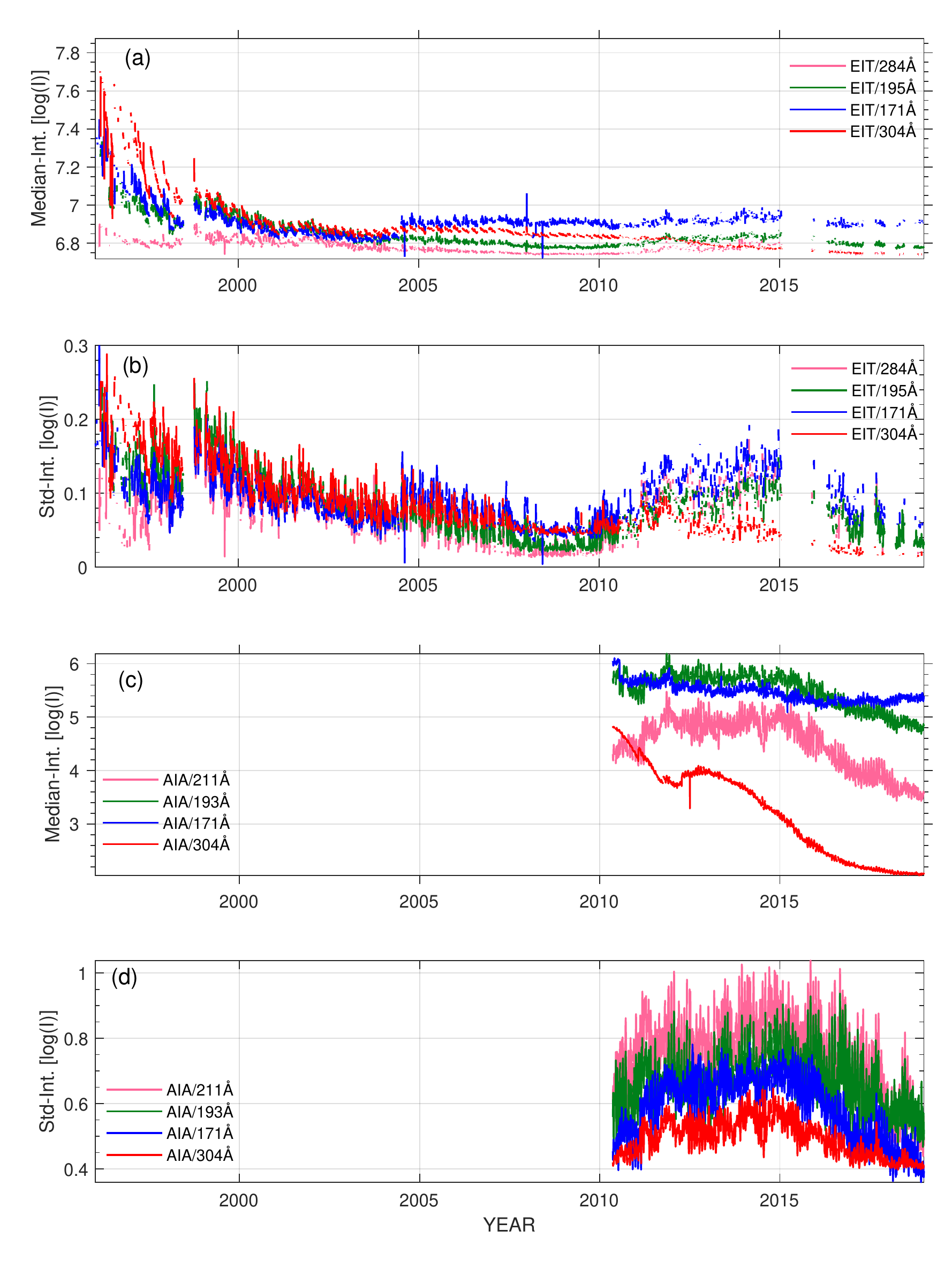}}
\caption{Median logarithmic intensities and standard deviations of SOHO/EIT (a, b) and SDO/AIA (c, d) full-disk imaged for the four different wavelengths.}\label{F4}
\end{figure}

Since June 19, 2003, due to a malfunction in the antenna pointing mechanism, SOHO's signals were relayed primarily on the backup antenna. Because of this the spacecraft is systematically rolled by 180$^\circ$ around the solar direction every three months in order to have the Earth constantly in the antenna's field of view as the spacecraft and the Earth progress in their respective orbits (Figure \ref{F3}a). After October 29, 2010, the spacecraft is no longer aligned with the solar rotation axis but to the Ecliptic North Pole. Therefore, the roll angle with respect to the solar rotation axis goes from -7.25$^\circ$ in June to 7.25$^\circ$ in December. Figure \ref{F3}b,c shows the variation of the logarithmic EIT and AIA exposure times (\MakeUppercase{EXPTIME}) at different spectral lines.\\

Figures \ref{F4}a-b show the median and standard deviation of logarithmic pixel intensities of SOHO/EIT full-disk images for wavelengths 284{\AA}, 195{\AA}, 171{\AA} and 304{\AA}. Only the median intensity of 284{\AA} shows a somewhat steady variation throughout the entire depicted time interval. On the other hand, median intensities of 195{\AA}, 171{\AA} and 304{\AA} decrease especially during the first five years (1996-2001). This decrease is likely caused by the degradation of EIT sensitivity. The most significant degradation is observed in 304{\AA} intensity which had the highest values in 1996 and decreased systematically until 2004, increased a bit in 2005 and declined thereafter until 2016. The standard deviations of the intensities of different wavelengths show a slightly more systematic temporal evolution than the median
intensities with lowest values around the solar minimum in 2008-2009, and a fairly clear variation over the solar cycle 23/24.\\

\subsection{SDO/AIA} 
  \label{S-sdo_aia}    

SDO/AIA images were downloaded from Joint SDO Operations Centre (JSOC), located at Stanford University. The FITS images at JSOC are based on level 1.5 data with a 2-minute cadence reduced to 2.4 arcsec pixels corresponding to 1024$\times$1024 pixel resolution. These images are produced approximately 7 days after the time of observation from level 1 definitive data. The keywords \MakeUppercase{QUALITY} and \MakeUppercase{EXP\_TIME} are used to check for image corruption (for examples see Figures \ref{F2}d-e) and to identify significantly overexposed or underexposed images (see, e.g., Figure \ref{F2}f), respectively. Note that the \MakeUppercase{QUALITY} keyword also indicates the enhanced noise caused by solar particles. \\

Median and standard deviation of SDO/AIA logarithmic intensities are shown in Figure \ref{F4}c-d. Again, a fairly systematic evolution over the solar cycle 24 is seen in the standard deviations (Figures \ref{F4}d) and partly also in the median intensities of 211{\AA} and 193{\AA} wavelengths (Figure \ref{F4}c) which correspond to the outer layers of the corona. The 304{\AA} intensity shows significant degradation, although intensity increased momentarily in September 2011 due to 
instrument bake-outs \citep{Boerner2014}. Degradation has been suggested to result from the accumulation of volatile contamination on the optics or detector telescopes \citep{Boerner2014}. Due to the differences in the instrument calibrations, degradation and operating orbit, the spectral line intensities of SOHO/EIT and SDO/AIA show a quite different temporal evolve over the overlapping time interval.

\section{EUV Synoptic maps} 
  \label{S-synop_maps}

\begin{figure} 
\centerline{\includegraphics[width=1.0\textwidth,clip=]{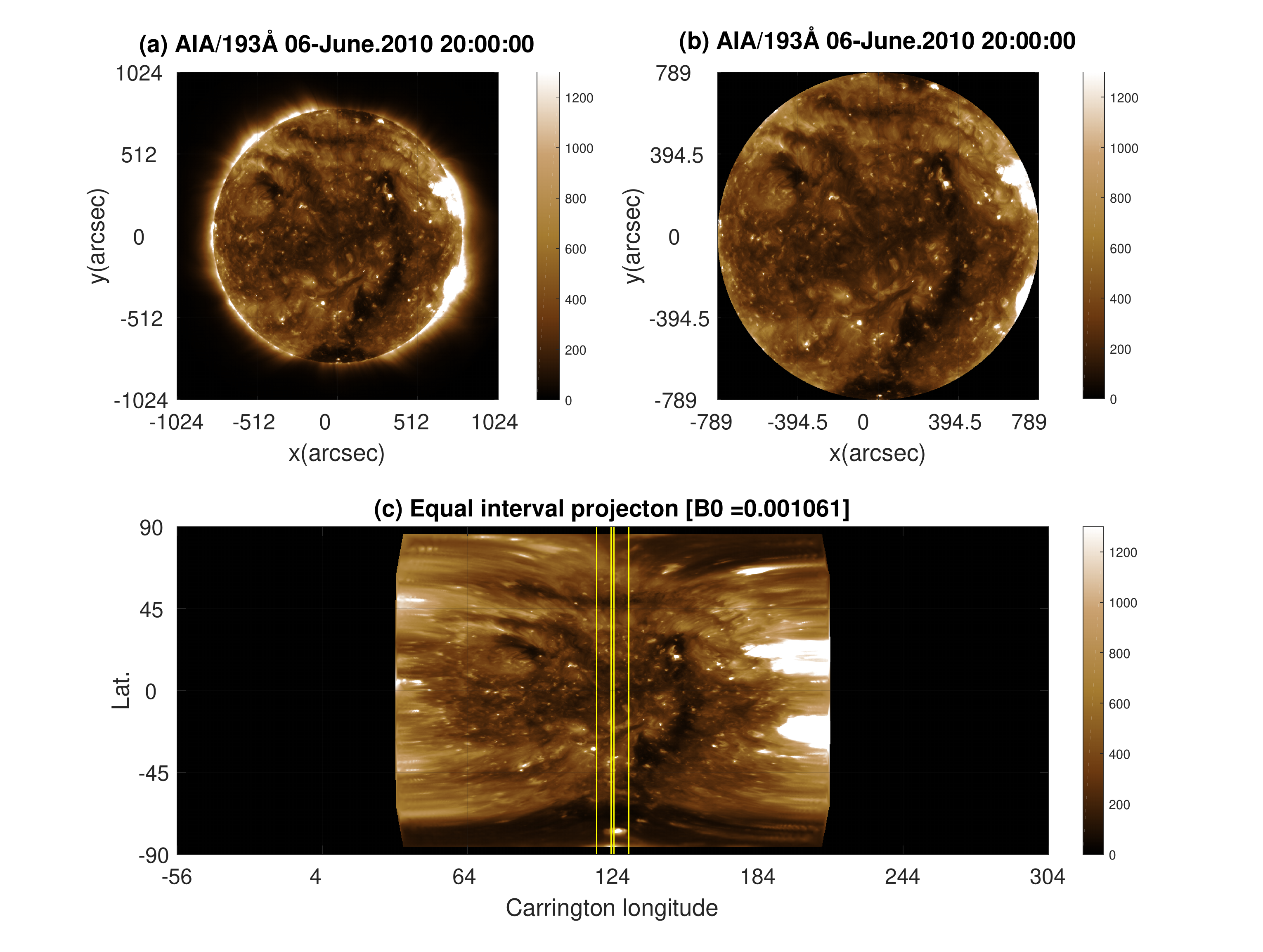}}
\caption{(a) Full-disk AIA image in 193{\AA} waelength taken on 06 June 2010. (b) The same with all off-limb pixels removed. (c) Full-disk image projected from image plane to heliographic coordinates. Vertical yellow dashed/solid lines comprise the wide($13.3^\circ$)/narrow($1.1^\circ$) central solar meridian strips, respectively.  }\label{F5}
\end{figure}

For each EUV full-disk image, we extract the information included within the image FITS header. This information includes the radius of the Sun in pixels (R\_SUN), Carrington longitude/latitude of the center of the solar disk (CRLN\_OBS/CRLT\_OBS) and Carrington rotation number (CAR\_ROT). The value of the solar photospheric radius (R\_SUN) is used to remove the off-limb measurements and to fix the image boundaries to the solar limb (see Figures \ref{F5}a-b).  Carrington latitude (CRLT\_OBS) of the disk center, the so-called $\beta_0$ angle, changes annually due to the tilt of the solar equator by $\pm7.25^\circ$ with respect to the Earth's ecliptic plane. This tilt also gives the Earth (or SOHO and SDO) a better view of the northern (southern) solar pole during fall (spring). The $\beta_0$ angle also affects the observed intensities in the following way: during fall (spring) the length of the line-of-sight path through the coronal matter to the southern (northern) hemisphere increases, which increases (decreases) the observed intensity of the EUV emission \citep{Hamada2018}. \\

After removing the off-limb pixels from full-disk images the visible solar disk is projected from image plane coordinates (X,Y) into heliographic latitude/longitude coordinates on the spherical solar surface ($\lambda$,$\phi$). In this projection the full-disk images are mapped to equal interval latitude-longitude grids (-90$^\circ$ to +90$^\circ$ latitude and $\pm 90^\circ$ longitude around the central meridian Carrington longitude) having 1440$\times$3600 pixels. The cartesian heliographic coordinates of an image plane point with pixel coordinates $X$ and $Y$ (these being zero at the center of the solar disk in the full-disk image) are 

\begin{eqnarray}
X_h &=& R_S\cos\lambda' \cos\phi'\cos\beta_0 + R_S\sin\lambda'\sin\beta_0, \\
Y_h &=& R_S\cos\lambda' \sin\phi',\\
Z_h &=& -R_S\cos\lambda'\cos\phi'\sin\beta_0 + R_S\sin\lambda'\cos\beta_0,
\end{eqnarray}
where $R_S$ is the solar radius in pixels and 

\begin{eqnarray}
\phi' &=& \arcsin\left(\frac{X}{\sqrt{R_S^2-Y^2}}\right) \\
\lambda' &=& \arcsin \left(\frac{Y}{R_S}\right).
\end{eqnarray}
The heliographic latitude ($\lambda$) and longitude ($\phi$) are then obtained from 

\begin{eqnarray}
\lambda &=& \arcsin\left(\frac{Z_h}{R_S}\right) \\
\label{eq_phi} \phi &=& \phi_0 + \arctan \left(\frac{Y_h}{X_h}\right),
\end{eqnarray}
where $\phi_0$ is the Carrington longitude of the central solar meridian (CRLT\_OBS).\\

The synoptic maps are constructed by concatenating meridional strips around the central solar meridian 
taken from the projected full-disk images. The image cadence of SOHO/EIT is typically several hours and
often the SOHO/EIT dataset contains gaps of several days. Because of this we selected 13.3$^\circ$ wide strips for
SOHO/EIT, which corresponds to 1 image/day (dashed vertical lines in Figure \ref{F5}c). For SDO/AIA the image cadence is higher. Thus, for SDO/AIA we produced two sets of synoptic maps, one using a central solar meridian strips 1.1$^\circ$ wide and another using 13.3$^\circ$ wide strips. The SDO/AIA maps with 13.3$^\circ$ strip width are thus constructed to be compatible with the SOHO/EIT maps that have the same strip width.\\

\begin{figure} 
\centerline{\includegraphics[width=1.0\textwidth,clip=]{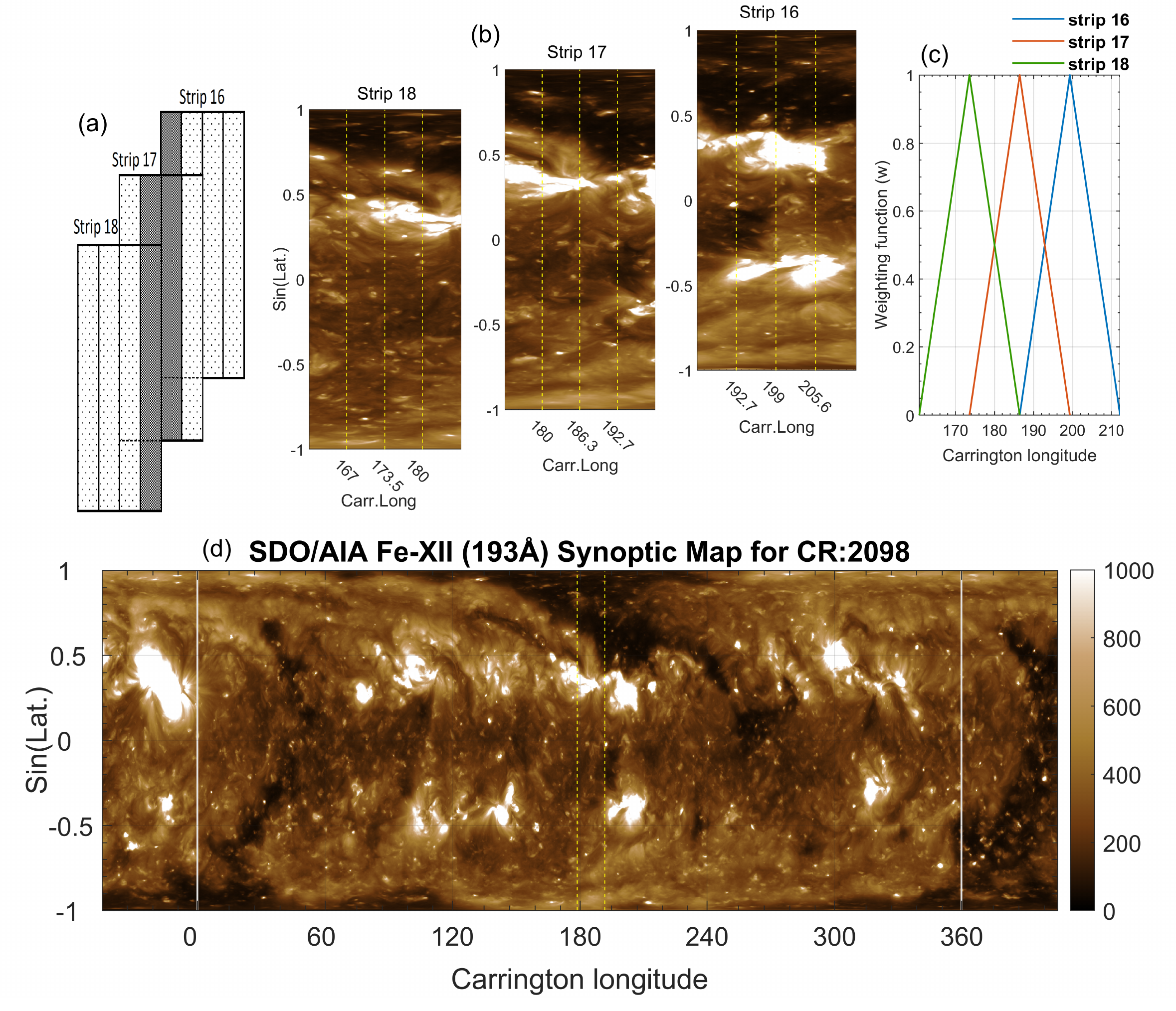}}
\caption{The synoptic map is made by averaging a central longitudinal strips taken from successive full-disk images projected to heliographic coordinates, with a half strip-width overlap between adjacent strips. (a) Three successive daily meridian strips 16, 17 and 18. The shaded area represents the overlapped parts. (b) The same EUV meridian strips where the shaded area located within the vertical yellow solid lines. (c) The corresponding linear weighting function for strips 16,17 and 18. (d) EUV synoptic map for CR 2098. Vertical white solid lines represent the map boundaries.}\label{F6}
\end{figure}

Figure \ref{F6} shows how the synoptic maps are constructed by including all the projected full-disk images
belonging to the corresponding Carrington rotation, taking from each the central meridian strip and overlapping
the neighbouring strips by half of their width. As an example see the central strip (no.17) in Figure \ref{F6} which is 
half overlapped with the previous and next strips 16 and 18 (Figure \ref{F6}a), respectively. The overlapped parts (see Figure \ref{F6}b), are averaged using a linearly varying weighting function (Figure \ref{F6}c) that drops from one at strip center to zero at the strip edge. The pixel intensity $S(\phi)$ at Carrington  longitude $\phi$ can be expressed as 

\begin{equation}
S(\phi) = \frac{\mathop{\sum}_{i=1}^{N} w_i(\phi)S_i(\phi)}{\mathop{\sum}_{i=1}^{N} w_i(\phi)},
\end{equation}
where $i$ loops through all the $N$ central meridian strips $S_i(\phi)$ taken from the projected full-disk images, including the corresponding longitude of $(\phi)$ of the same Carrington rotation. The weighting functions of each strip are defined as

\begin{eqnarray}
w_i(\phi) &=& 1 - 2\frac{\left| \phi - \phi_{0,i} \right|}{\Delta\phi},~\textnormal{when}~|\phi-\phi_{0,i}| < \frac{\Delta\phi}{2} \\
w_i(\phi) &=& 0,~\textnormal{when}~|\phi-\phi_{0,i}| \geq \frac{\Delta\phi}{2},
\end{eqnarray}
where $\phi_{0,i}$ is the central Carrington longitude of the $i$:th strip (see also Eq. \ref{eq_phi}) and
$\Delta\phi$ is the width of the strip in degrees.\\

\begin{figure} 
\centerline{\includegraphics[width=1.0\textwidth,clip=]{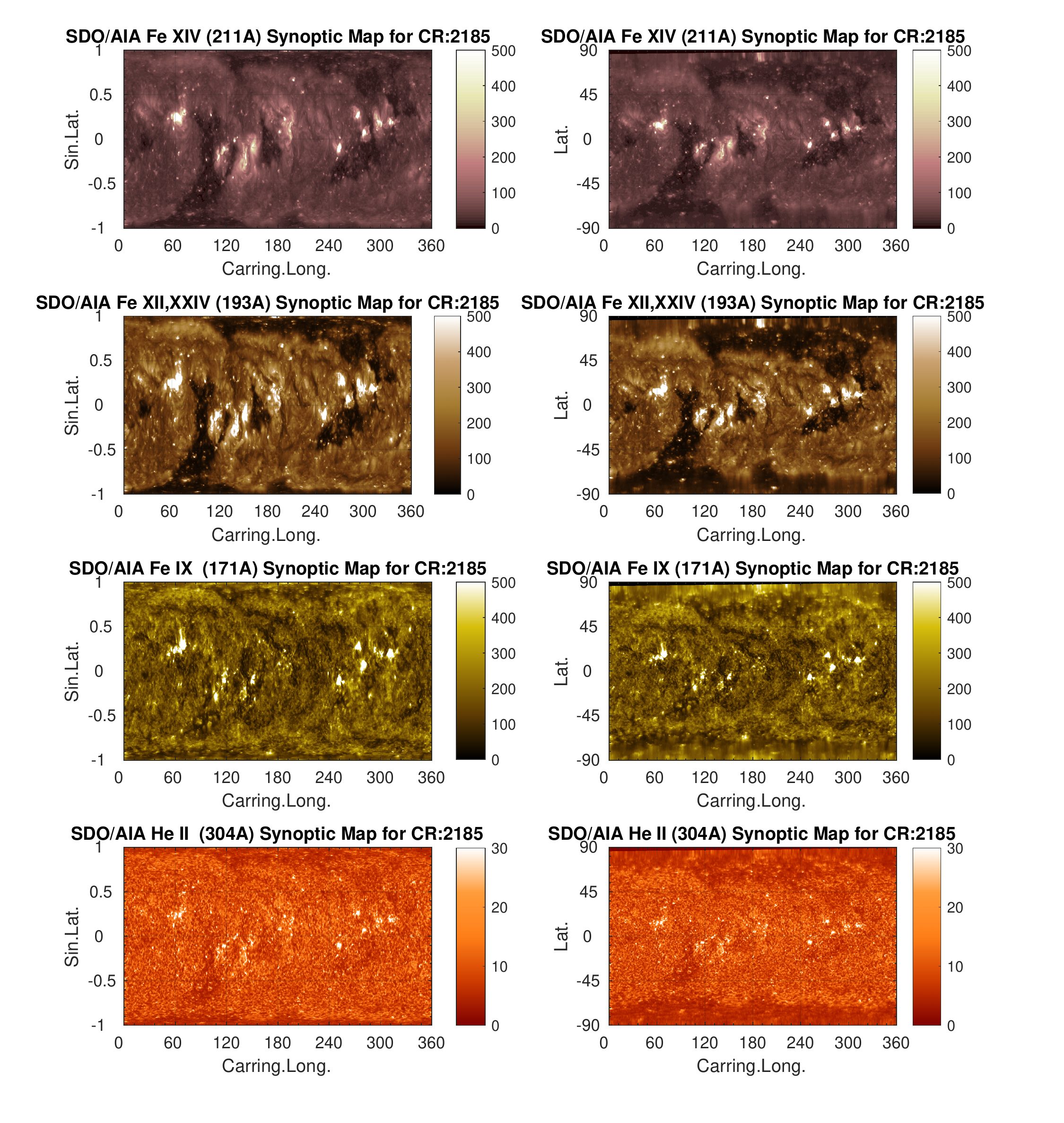}}
\caption{SDO/AIA synoptic maps for CR 2185 (2016 Dec. 14) for the four wavelengths (211{\AA}, 193{\AA}, 171{\AA} and 304{\AA}). Solar synoptic maps are presented in sine (left charts) and linear (right charts) latitude.}\label{F7}
\end{figure}

Sometimes there are relatively long gaps in the AIA or EIT datasets, which would result in gaps in the synoptic maps if the strip width remained constant. To fill in such gaps we extend the strips equally on both sides of the data gap up to a maximum of 3 days (i.e., about 40.89$^\circ$). The strip width is limited because extending the strips too much would start introducing significant distortions due to projection effects (see Figure \ref{F5}c). All the final synoptic maps have a resolution of 1440$\times$3600 pixels, covering all measured latitudes from -90$^\circ$ to +90$^\circ$ and longitudes from 0$^\circ$ to 360$^\circ$. Note however, that because of visibility limitation related to the annual $\beta_0$-angle variation there is an annually varying data gap at polar latitudes. For each Carrington rotation, the maps are produced both in linear and sine latitude. Figure \ref{F7} shows an example of the constructed synoptic maps of CR 2185 in sine and linear latitude for SDO/AIA 211{\AA}, 193{\AA}, 171{\AA}, and 304{\AA}. For SDO/AIA, EUV synoptic maps were constructed from Carrington rotation CR 2097 (2010 May 20) to CR 2204 (2018 May 16) while for SOHO/EIT the maps were constructed from CR 1906 (1996 Feb 19) to CR 2212 (2018 Dec 20) (Table \ref{T4}). \\

\begin{table}
\caption{Data coverage of the new SOHO/EIT and SDO/AIA EUV synoptic maps}
\label{T4}
\begin{tabular}{ccccc}     
\hline               
Observable & \multicolumn{2}{c}{Start} & \multicolumn{2}{c}{End} \\
    & CR & Date & CR & Date\\
\hline               
SOHO/EIT: 284{\AA}, 195{\AA}, 171{\AA} and 304{\AA}  & 1906 & ~1996.02.13 & 2212 &  ~2018.12.20 \\
SDO/AIA: 211{\AA}, 193{\AA}, 171{\AA} and 304{\AA}    &  2097 &  ~2010.05.20 &  2212 &  ~2018.12.20\\            
\hline               
\end{tabular}
\end{table}

\begin{figure} 
\centerline{\includegraphics[width=1.0\textwidth,clip=]{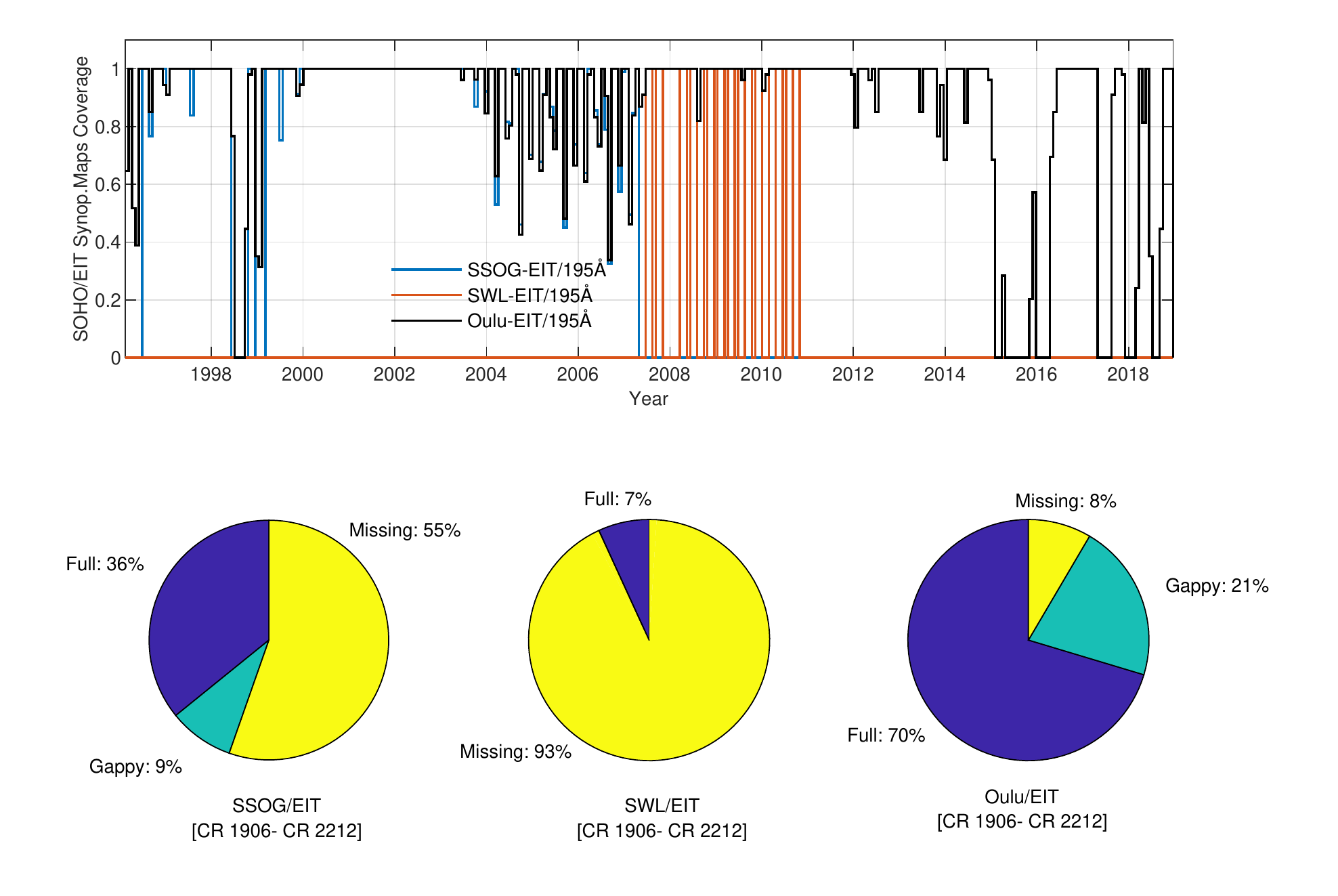}}
\caption{Data coverage of all (full and gappy) SOHO/EIT synoptic maps for 195{\AA}  from SSOG (Standford University), SWL (George Mason University) and the new maps produced here. The top panel shows the data coverage fraction as a function of time for the three datasets. The pie charts in the bottom display the overall percentages of full, gappy and missing synoptic maps for the three datasets.}\label{F8}
\end{figure}

\begin{figure} 
\centerline{\includegraphics[width=1.0\textwidth,clip=]{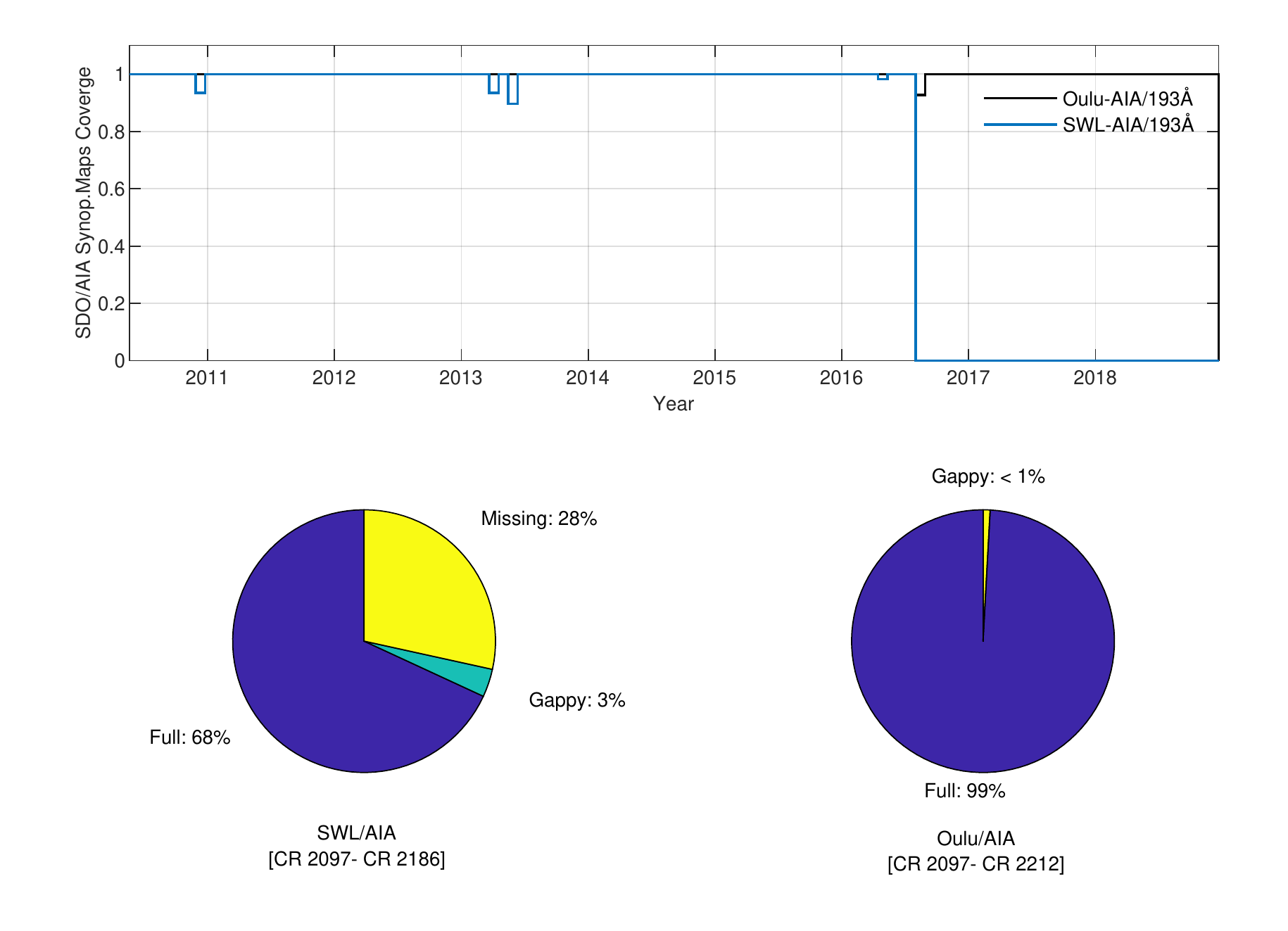}}
\caption{Data coverge of all (full and gappy) SDO/AIA synoptic maps for 193{\AA} from SWL and from the new synoptic maps presented here. The pie chart represent the percentage of full, gappy and missing synoptic maps for the two datasets.}\label{F9}
\end{figure}

In comparison to the existing SOHO/EIT and SDO/AIA synoptic map datasets from SSOG and SWL, the new maps constructed here provide a longer and homogeneous dataset with fewer data gaps. To demonstrate this, Figures \ref{F8} and \ref{F9} show EIT/195{\AA} and AIA/193{\AA} data coverages as a pie chart representing the overall fraction of full, gappy and missing synoptic maps, respectively. Overall the SOHO/EIT full-disk image dataset available from February 1996 until December 2018 covers 307 Carrington rotations. SWL provides only 21 full synoptic maps for SOHO/EIT, and SSOG 110 full maps. In contrast, our new dataset contains a total of 216 full maps for this period and, thus, considerably extends the coverage compared to the SSOG and SWL datasets. This number of full maps includes those gappy ones filled up to 3 missing days.\\

For SDO/AIA, the full-disk images are available from CR 2097 (May 2010) until present. The SDO/AIA synoptic maps offered by SWL extend until CR 2186 with a total of 79 full maps (68\%) and 4 gappy maps (3\%), respectively, while our new dataset is almost complete until CR 2215, with 118 (99\%) of Carrington rotations are fully covered and only one map (less than 1\%) containing gappy synoptic maps.

\section{Homogenization of SOHO/EIT and SDO/AIA synoptic maps} 
      \label{S-eit_aia} 

\begin{figure} 
\centerline{\includegraphics[width=1.0\textwidth]{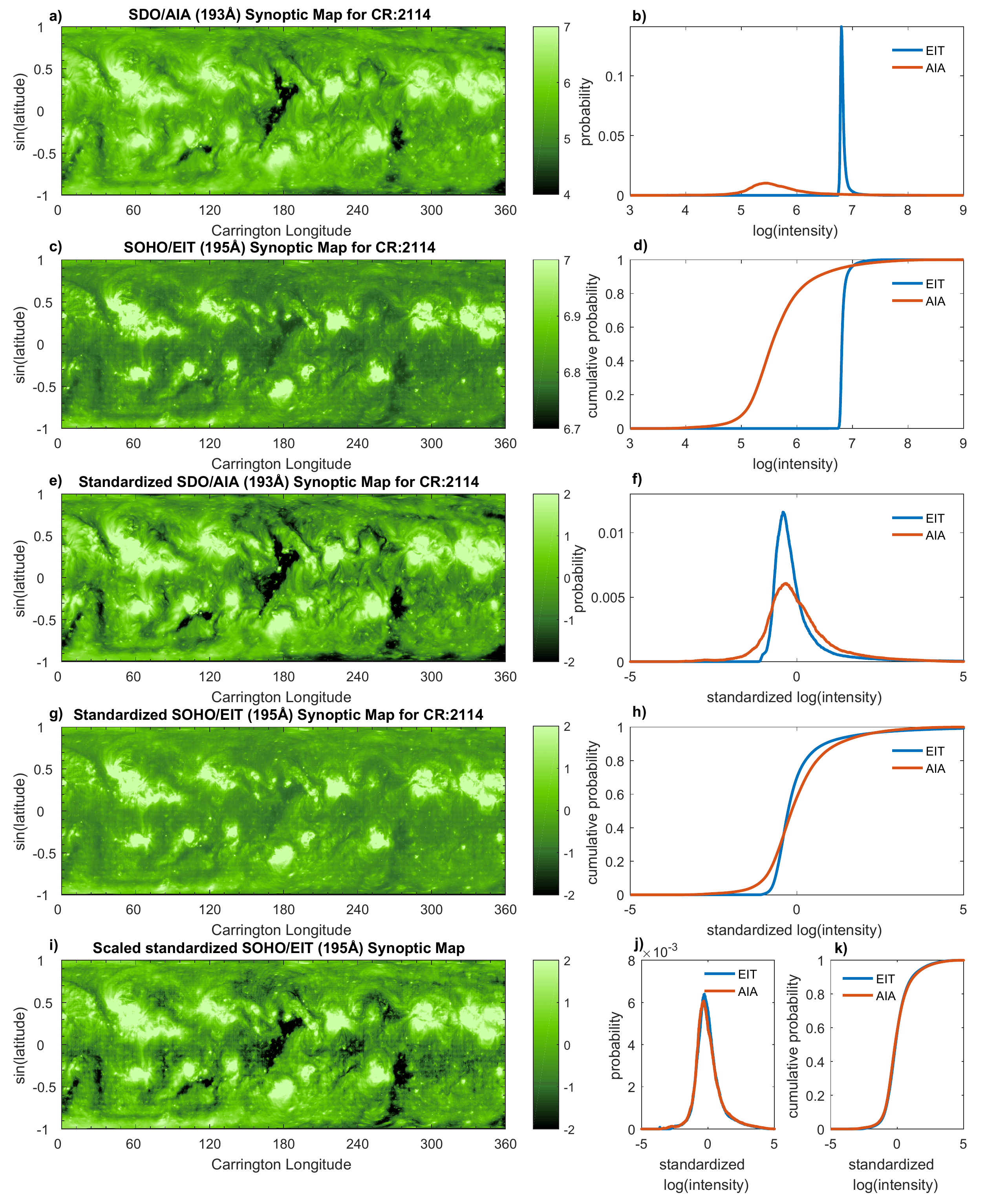}}
\caption{Homogenization of synoptic maps for CR 2114 (starting 26 Aug 2011) for SDO/AIA 193\AA\ and SOHO/EIT 195\AA. 
Panels a), c), e), g), i) on the left depict SDO/AIA, SOHO/EIT, standardized SDO/AIA, standardized SOHO/EIT and scaled standardized SOHO/EIT
synoptic maps respectively. Panels b), d) show the histograms and cumulative histograms, respectively, of the synoptic maps in a) and c). 
Panels f), h) show the histograms and cumulative histograms, respectively, of the standardized synoptic maps in e) and g). 
Panels j) and k) show the comparison between the histograms of scaled standardized SOHO/EIT and standardized SDO/AIA corresponding to 
the synoptic maps in e) and i). All the intensities are in logarithmic scale. In the histogram plots the blue curves correspond to SOHO/EIT
and the red curves to SDO/AIA.}\label{F10}
\end{figure}

Because of instrumental differences the overall intensity distributions of SOHO/EIT and SDO/AIA images of corresponding wavelengths are very different. As an example of this, Figure \ref{F10} shows AIA 193\AA\ (Fig. \ref{F10}a) and EIT 195\AA\ (Fig. \ref{F10}c) synoptic maps for CR 2114 in logarithmic intensity scale. Panels b and d in Figure \ref{F10} display the histogram and cumulative histogram of their logarithmic intensities. One can see a dramatic difference between AIA and EIT in the form of the intensity distributions (Fig. \ref{F10}b) and the range of intensities they cover. It is clear that in order to have a homogeneous composite dataset combining the two instruments, the synoptic maps from the two instruments need to be inter-calibrated to the same level. Here we choose to transform the EIT maps to AIA level. 

Before inter-calibration all the EIT and AIA synoptic map data are converted to logarithmic scale and then standardized separately for each map by subtracting the average logarithmic intensity of the map and dividing by the standard deviation of logarithmic intensity of the corresponding map. 
We used logarithmic intensities because in linear scale the intensity distributions are extremely skewed. In logarithmic scale the histograms are much more Gaussian and the distribution mean and standard deviation better represent the center and spread of values. The smoother and more symmetric
histograms also allow a more accurate determination of cumulative probabilities for each pixel value, which is important for the homogenization as will be discussed below. Standardization scales the synoptic map histograms to the same level and thereby, e.g., removes the intensity changes related to instrument degradation.
However, these changes are very difficult to separate from real, e.g., solar cycle related changes. Standardized synoptic maps describe relative intensity variations within each map, and thus do not have this problem. Note that, these synoptic maps cannot be used to study the long-term changes of the average EUV intensities, only the relative intensity variations and relative spatial distribution changes. Figures \ref{F10}e and \ref{F10}g show the standardized SDO/AIA and SOHO/EIT synoptic maps for CR 2114, while \ref{F10}f and \ref{F10}h show the corresponding histograms and cumulative histograms. 
One can see that the standardization of logarithmic intensities improves the agreement between the EIT and AIA maps and corresponding histograms. However, significant differences still remain, which is seen in the different form of the two histograms. The EIT--AIA inter-calibration aims to eliminate these differences in the form of the histograms by finding a transformation that brings the SOHO/EIT histogram as close to the corresponding SDO/AIA histogram as possible.

The inter-calibration is based on comparing 93 simultaneous full synoptic maps from EIT and AIA between CR 2097- CR 2212. For all these maps we first compute the histogram of standardized logarithmic intensity and then compute the overall average histograms for those 93 CRs, separately for SOHO/EIT and SDO/AIA. We then form the corresponding overall cumulative histograms. Using these two cumulative histograms we compute 
for each value of SOHO/EIT standardized log-intensity $I_{\textnormal{EIT}}$ the corresponding SDO/AIA standardized log-intensity 
$I_{\textnormal{AIA}}$, which has the same cumulative probability value in its respective histogram as the SOHO/EIT value. 
The difference of these values, i.e., $I_{\textnormal{diff}}(I_{\textnormal{EIT}}) = I_{\textnormal{AIA}}-I_{\textnormal{EIT}}$ 
is a function of the SOHO/EIT standardized log-intensity and indicates by how much the standardized log-intensities of SOHO/EIT
pixels should be changed in order for the pixel value to correspond to the AIA level.
Thus, the scaled standardized log-intensities of SOHO/EIT pixels are obtained by equation

\begin{equation}
I_{\textnormal{EIT,scaled}} = I_{\textnormal{EIT}} + I_{\textnormal{diff}}(I_{\textnormal{EIT}}).
\end{equation}

\begin{figure} 
\centerline{\includegraphics[width=1.0\textwidth]{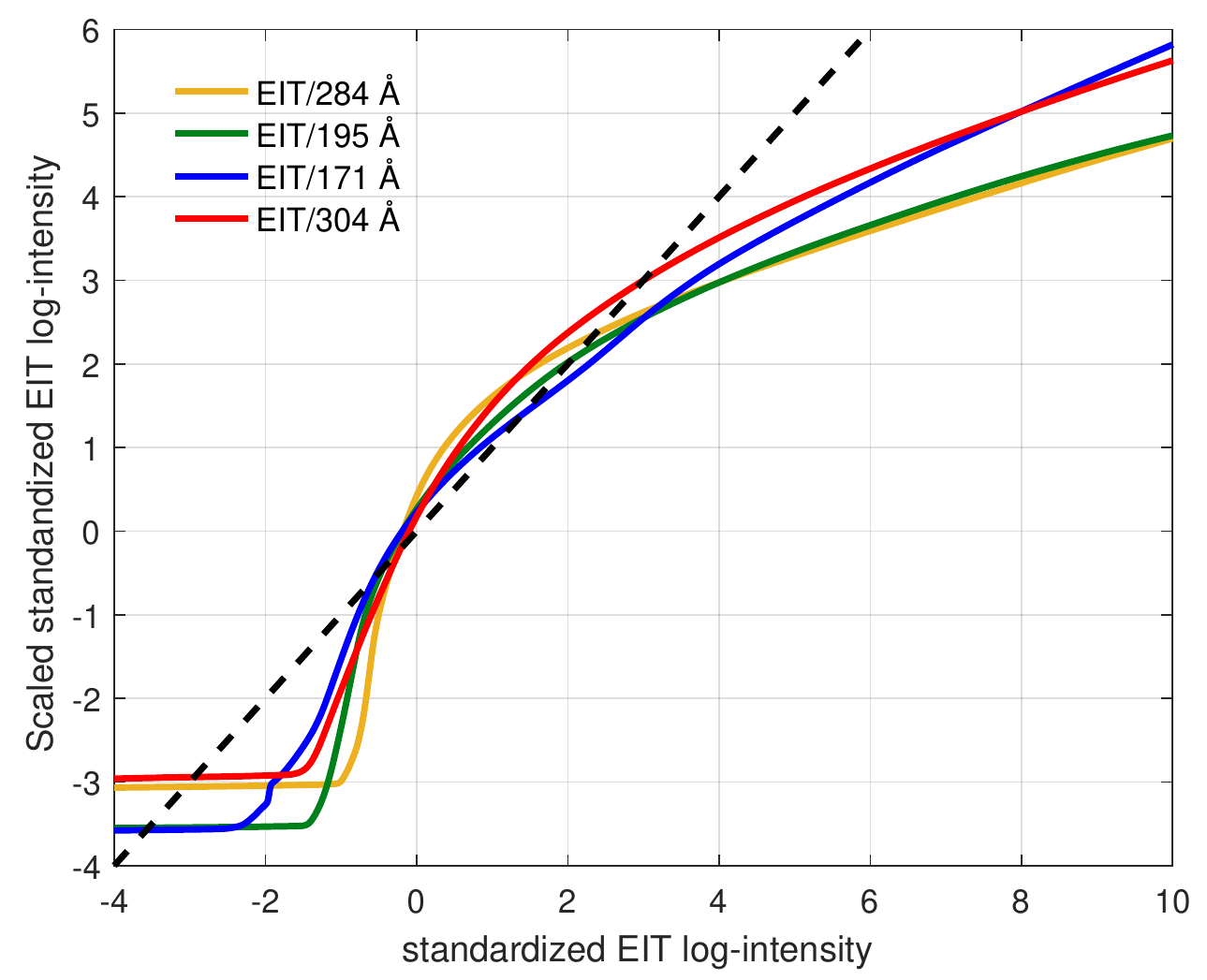}}
\caption{Relationships between the scaled standarized EIT log-intensities and the original standardized EIT log-intensities.
The different colors correspond to the four different EIT wavelengths as indicated. The black dashed line shows the one-to-one 
relationship for reference.}\label{F11}
\end{figure}

This scaling inter-calibration procedure is performed for each synoptic map of the four wavelengths separately.
The same procedure is applied to all EIT maps including those ones that were not used to determine the EIT/AIA inter-calibration.
The relationships between the scaled and original standardized log-intensities for the four EIT wavelengths are shown in 
Figure \ref{F11}. The figure also indicates the one-to-one line as a reference. One can see that for 
all four wavelengths the intensities between -3 and 0 (i.e., up to 3 standard deviations below the mean) are decreased
from the original value by scaling. Intensities between about 0 and 3 are (i.e., up to 3 standard
deviations above the mean intensity) are increased by the scaling. Finally the very dark values ($<$-4) are increased 
and the very bright values ($>$4) are decreased by scaling. 

Figure \ref{F10}i shows the scaled standardized SOHO/EIT synoptic map, which should be compared with the standardized SDO/AIA 
synoptic map shown in Figure \ref{F10}e. Visual comparison between these two synoptic maps shows that they are very similar.
The contrast and range of standardized intensity values are roughly the same and, e.g., dark coronal holes seem to be resolved quite similarly in both maps. Figures \ref{F10}j and \ref{F10}k show the histograms and cumulative histograms, respectively, of the scaled standardized SOHO/EIT map and standardized SDO/AIA map. One can see that the scaling (inter-calibration) procedure discussed above efficiently transforms the EIT histogram to closely match with the corresponding AIA histogram, thereby validating the inter-calibration for CR 2114. The same is true also for all the other maps used in the inter-calibration.

\begin{figure} 
\centerline{\includegraphics[width=1.0\textwidth]{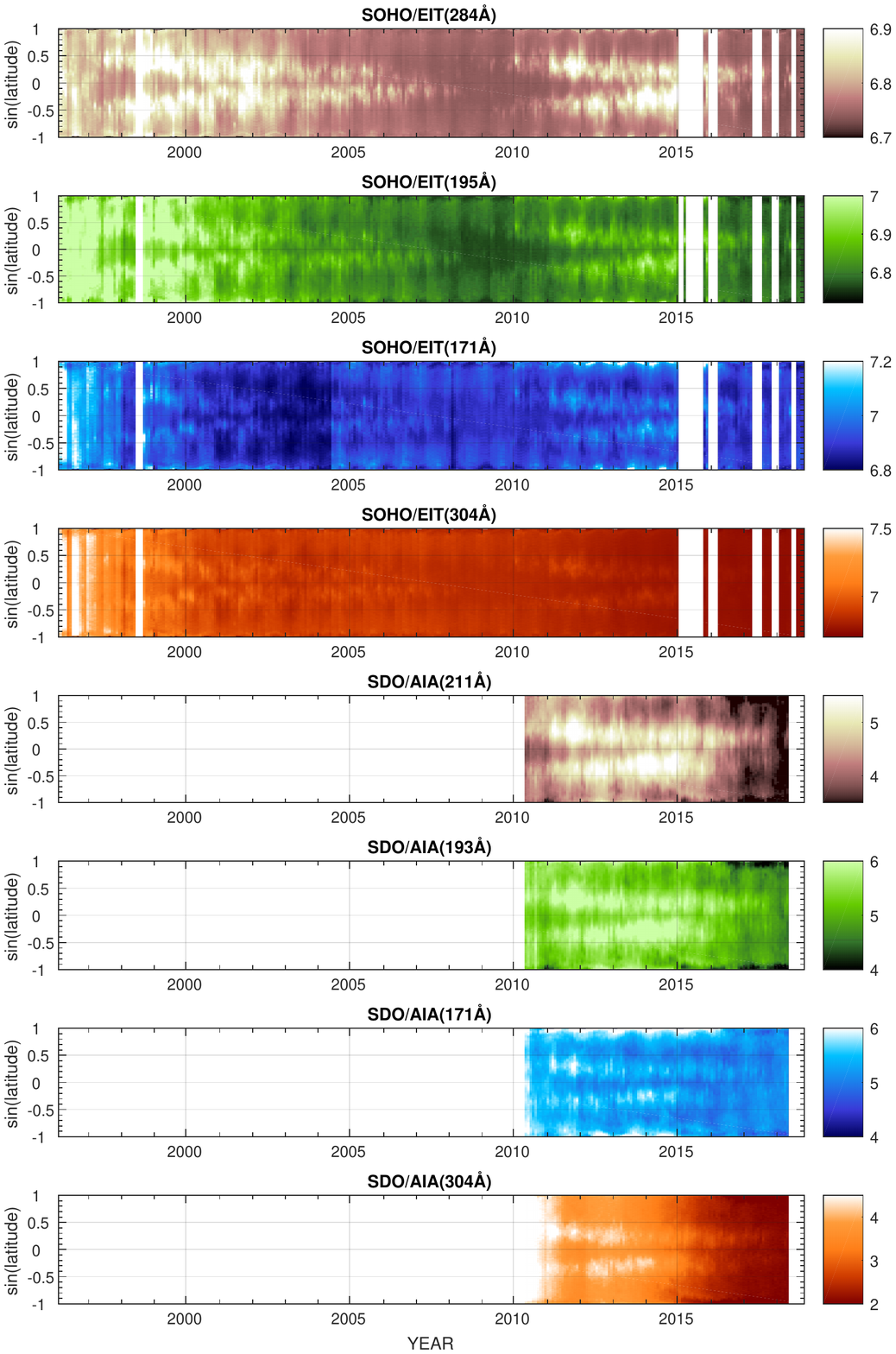}}
\caption{Distribution of original logarithmic EUV intensities as a function of heliographic sine latitude and time. 
The distributions have been averaged over Carrington longitude. The four top rows correspond to SOHO/EIT 
intensities in wavelengths 211{\AA}, 195{\AA}, 171{\AA} and 304{\AA}. The four bottom rows correspond to the SDO/AIA
intensities in wavelengths 211{\AA}, 193{\AA}, 171{\AA} and 304{\AA}.}
\label{F12}
\end{figure}

\begin{figure} 
\centerline{\includegraphics[width=1.0\textwidth]{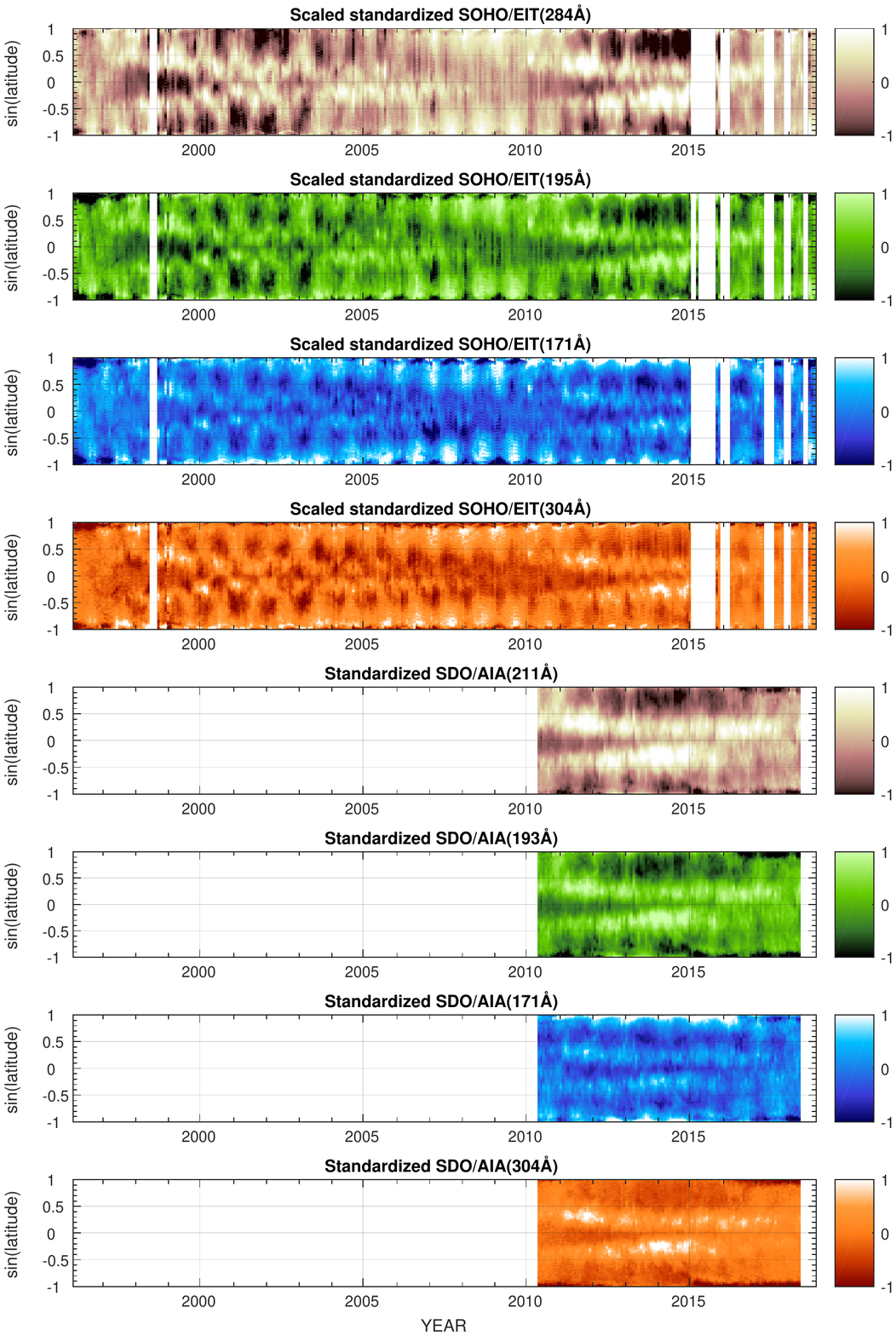}}
\caption{Same as Figure \ref{F12}, but with SOHO/EIT standardized and scaled and SDO/AIA standardized.}
\label{F13}
\end{figure}

Figure \ref{F12} shows a broad overview of logarithmic EUV intensities of the SOHO/EIT (first four rows) and SDO/AIA (bottom four rows) synoptic maps over the entire time interval covered by the maps. The intensity distributions in Fig. \ref{F12} are obtained from the synoptic maps by averaging over Carrington longitude. One can see the previously discussed drifts  in the SOHO/EIT intensities in all the four wavelengths (see Figure \ref{F4}a,b and related discussion). 
For SDO/AIA such a drift is especially visible in the 304{\AA} wavelength. In addition, the 171{\AA} wavelength of SOHO/EIT displays an abrupt change in the intensity distribution in 2004, which was also seen in Figure \ref{F4}c,d. \\

As a comparison to Figure \ref{F12}, Figure \ref{F13} shows the corresponding relative EUV intensities obtained from the final homogenized SOHO/EIT and SDO/AIA synoptic maps, i.e., with scaled standardized log-intensities for EIT and standardized log-intensities for AIA.
Here one can see that standardization effectively removes the above discussed temporal drifts in the overall intensity, from both datasets. It should be noted that standardization also removes the real solar cycle related variation of intensities. However, for the main purpose of producing these relative intensity maps, i.e., the determination of coronal holes, this is not a problem. The scaling of SOHO/EIT maps brings them to the same relative intensity level as the corresponding AIA maps. Visual inspection of the corresponding wavelengths in Fig. \ref{F13} shows that the two datasets produce indeed very similar distributions for the corresponding wavelengths.

\section{Summary} 
      \label{S-Summary} 

In this study we presented a new dataset of EUV synoptic maps based on SOHO/EIT and SDO/AIA full disk images. The new dataset differs from the previously existing datasets in several ways. First of all, the new synoptic maps have been constructed using the same spatial resolution (13.3$^\circ$ wide central solar meridian strip) for both instruments thereby reducing the differences between the maps that may result from different image resolution or different image sampling frequency. We have also used the entire set of full-disk images available at the time of writing and thereby notably extended the coverage of synoptic maps compared to the previous datasets. 
Taking advantage of the higher image cadence in SDO/AIA we also constructed another set of SDO/AIA synoptic maps with narrower central meridian strips (1.1$^\circ$ wide) 
which allow each synoptic map longitude to better correspond to the central solar meridian.
Secondly, in order to avoid the problems due to significant and complicated drifts in the SOHO/EIT and SDO/AIA intensities, we standardized the synoptic maps in logarithmic intensity scale. While this procedure inevitably removes the real, e.g., solar cycle related intensity variations it produces a more homogeneous dataset of relative intensities for the purpose of studying the evolution of coronal structures, e.g., coronal holes. Thirdly, even after standardizing the synoptic maps the relative 
intensity distributions of the corresponding SOHO/EIT and SDO/AIA synoptic maps are very different due to instrumental differences. In order to remove these differences between the EIT and AIA maps we developed a method to scale the EIT pixel values to the AIA level.
The scaling is based on comparing the overall average cumulative histograms of 93 simultaneous EIT and AIA synoptic maps and determining how much the EIT standardized
log-intensities should be changed so that their overall cumulative probability matches that of AIA.
As a result of this scaling the simultaneous EIT and AIA synoptic maps are very close to each other.
Overall, this work resulted in a new homogenized and standardized database of solar EUV synoptic maps from SOHO/EIT and SDO/AIA.
The constructed maps describe relative logarithmic intensity variations within each map. These maps are well suited for studying the evolution of 
different coronal structures such as coronal holes over solar cycle timescales, where data homogeneity is important.
\begin{acks} 
The dataset constructed here is available by request to the authors (A. Hamada, amr.hamada@oulu.fi; T. Asikainen, timo.asikainen@oulu.fi or K. Mursula, kalevi.mursula@oulu.fi).
We acknowledge the financial support by the Academy of Finland to the ReSoLVE Center of Excellence (project No. 307411).
\end{acks}


\bibliographystyle{spr-mp-sola}

\IfFileExists{\jobname.bbl}{} {\typeout{}
\typeout{****************************************************}
\typeout{****************************************************}
\typeout{** Please run "bibtex \jobname" to obtain} \typeout{**
the bibliography and then re-run LaTeX} \typeout{** twice to fix
the references !}
\typeout{****************************************************}
\typeout{****************************************************}
\typeout{}}

\end{article} 

\end{document}